\documentclass[10pt]{article}
\usepackage[a4paper,top=1.8cm,bottom=1.9cm,left=1.7cm,right=1.7cm]{geometry}
\usepackage[T1]{fontenc}
\usepackage{lmodern,amsmath,amssymb,bm,graphicx,booktabs,microtype}
\usepackage[numbers,sort&compress]{natbib}
\usepackage[font=small,labelfont=bf,labelsep=period]{caption}
\usepackage{titlesec}
\usepackage[colorlinks=true,citecolor=black,linkcolor=black,urlcolor=black]{hyperref}
\titleformat{\section}{\normalfont\normalsize\bfseries}{\thesection.}{0.5em}{}
\titleformat{\subsection}{\normalfont\normalsize\itshape}{\thesubsection.}{0.5em}{}
\titlespacing*{\section}{0pt}{2.0ex plus .5ex minus .2ex}{0.9ex plus .2ex}
\titlespacing*{\subsection}{0pt}{1.4ex plus .3ex minus .2ex}{0.6ex plus .2ex}
\graphicspath{{figures/}}
\hypersetup{pdftitle={What Does Precision Spectroscopy Actually Constrain about $X_{17}$ Couplings?},pdfauthor={Raoul Serao and Antonio Capolupo},pdfsubject={What spectroscopy constrains: coupling products, identifiability and individual bounds},pdfkeywords={$X_{17}$, isotope shifts, muonic atoms, anomalous magnetic moments}}
\newcommand{\epse}{\epsilon_e}
\newcommand{\epsm}{\epsilon_\mu}
\newcommand{\epsp}{\epsilon_p}
\newcommand{\epsn}{\epsilon_n}
\newcommand{\mx}{m_X}
\newcommand{\dd}{\mathrm{d}}
\newcommand{\fm}{\mathrm{fm}}
\newcommand{\mev}{\mathrm{meV}}
\newcommand{\MeV}{\mathrm{MeV}}
\begin{document}
\setlength{\abovedisplayskip}{6pt plus 2pt minus 1pt}
\setlength{\belowdisplayskip}{6pt plus 2pt minus 1pt}
\setlength{\abovedisplayshortskip}{3pt plus 1pt}
\setlength{\belowdisplayshortskip}{4pt plus 1pt minus 1pt}
\begin{center}
{\Large\bfseries What Does Precision Spectroscopy\\[2pt]Actually Constrain about $X_{17}$ Couplings?\par}
\vspace{7pt}
{\normalsize Raoul Serao$^{1,2,*}$ and Antonio Capolupo$^{3,*}$\par}
\vspace{4pt}
{\small $^1$Scuola Superiore Meridionale,
Largo San Marcellino 10, 80138 Napoli, Italy;\\[2pt]
$^2$Istituto Nazionale di Fisica Nucleare,
Sezione di Napoli, Complesso Universitario
di Monte S. Angelo, Via Cintia Edificio 6,
80126 Napoli, Italy\\[2pt]
$^3$Dipartimento di Fisica E.~R. Caianiello, Universit\`a di Salerno, and INFN -- Gruppo Collegato di Salerno,\\
Via Giovanni Paolo II 132, 84084 Fisciano (SA), Italy\\[2pt]
$^*$Correspondence: r.serao@ssmeridionale.it, capolupo@sa.infn.it\par}
\end{center}
\begin{abstract}
Can precision spectroscopy disentangle the individual couplings of the $X_{17}$ boson? Within a leading one-body model, we combine electronic and muonic spectroscopy of hydrogen, deuterium and helium with lepton magnetic moments. We show that hydrogen–deuterium comparisons can lose neutron sensitivity through cancellations, while the combined observables restore generic local identifiability without guaranteeing finite individual bounds. The unrestricted fit admits formally unbounded nucleon-coupling directions, some of which violate the assumptions underlying the physical model. Finite-coupling tests nevertheless reveal statistically compatible solutions, with bounds dependent on nuclear responses and physical assumptions. Our results establish what precision measurements genuinely constrain and what additional information is required to extract meaningful limits on $X_{17}$ couplings.
\end{abstract}
\vspace{0.2em}

\section{Introduction}
\label{sec:intro}
Precision atomic spectroscopy tests new interactions through small shifts of transition frequencies \cite{Jaeckel2010,Karshenboim2010}. Electronic and muonic atoms probe different spatial scales because their bound leptons have different masses. Isotope comparisons change the nuclear proton--neutron content while retaining a common electronic structure. They therefore add sensitivity to neutron couplings \cite{Delaunay2017,Berengut2018}.

$X_{17}$ provides a concrete application. This proposed boson was introduced to explain electron--positron excesses in beryllium \cite{Krasznahorkay2016}, followed by anomalies in helium and carbon \cite{Krasznahorkay2021,Krasznahorkay2022}. Vector interpretations allow distinct electron, muon, proton and neutron couplings \cite{Feng2016,Feng2017}; Ref.~\cite{SeraoReview2026} reviews their phenomenology. Nuclear-transition and production searches test this hypothesis independently \cite{MEGII2025,PADME2025}. Spectroscopy instead probes virtual exchange and the resulting bound-state energy shifts.

Interpreting these shifts requires care. Spectroscopy measures frequencies, whereas nuclear radii are inferred using a theoretical description. Muonic hydrogen established the proton-radius puzzle \cite{Pohl2010,Antognini2013}, subsequently tested by electronic measurements \cite{Beyer2017,Bezginov2019,Grinin2020}. A short-range interaction can mimic a nuclear-size contribution \cite{Jentschura2018}. Radii extracted within the Standard Model (SM) can thus absorb part of the new signal. A comparison of electronic and muonic radii must propagate the interaction through both extractions.

Existing methods address different pieces of the problem. Potvliege and collaborators analyze hydrogen--deuterium and helium spectra and derive constraints on lepton--nucleon coupling products \cite{Potvliege2023,Potvliege2025}. Effective-field-theory treatments connect atomic responses to mediator mass and spin \cite{Frugiuele2022}, while finite-source calculations refine muonic responses \cite{Huang2026,Lin2026}. Delaunay and collaborators emphasize that adjusted constants and new interactions should be inferred together rather than treated as independent inputs \cite{Delaunay2023,DelaunayErratum2025}. These studies provide the spectroscopic framework; they should not be confused with a demonstration that every individual coupling can be bounded. Previous $X_{17}$ phenomenology combined magnetic moments with muonic Lamb shifts \cite{CapolupoX172025,SeraoTensions2026}, and separate electron and muon couplings were investigated in Ref.~\cite{SeraoG22026}.

Complementary precision strategies have also been proposed for nonlocal electromagnetic effects \cite{CapolupoNonlocalG22022}, dark-sector geometric phases \cite{SeraoGeometricDark2026}, scalar-field interferometry \cite{SeraoScalarInterferometry2026}, photon--axion phases \cite{CapolupoPhotonPhase2015}, and axion-mediated neutron forces \cite{CapolupoNeutron2021,CapolupoAxionEntanglement2020}. These involve different operators and are not imported as likelihood constraints here.

The remaining inference question is precise: which combinations does spectroscopy constrain, and when do they imply finite bounds on individual $X_{17}$couplings? Products alone do not fix a lepton normalization, and the proton and neutron contributions can cancel in certain response directions. Additional observables may remove an infinitesimal degeneracy without excluding distant parameter sequences. In particular, full rank of a local sensitivity matrix is not a statement about the size of a global confidence region. Conversely, a mathematical sequence that requires a large change of the nuclear binding potential cannot automatically be interpreted within an atomic model that keeps its nuclear sources fixed.

Here we retain four independent signed vector couplings, derive electronic and muonic radius responses, and combine two independent isotope comparisons with both lepton anomalous magnetic moments. Relative to existing light-atom analyses, our focus is explicitly the rank and global geometry of the four-parameter inference problem, including the distinction between removable local degeneracies, unbounded formal profiles, and nuclear-source validity. We additionally evaluate finite-coupling profiles under labelled nuclear-potential stress thresholds rather than presenting those thresholds as experimental exclusions. This distinction is essential because the nuclear many-body response to the new interaction has not been fitted. We use published radius summaries and do not claim to replace a global fit of raw transition frequencies, adjusted constants and nuclear theory errors.

The paper is organized as follows.
Section~\ref{sec:response} introduces the physical model
and derives the relevant spectroscopic and magnetic-moment
responses. Section~\ref{sec:identifiability} presents the
numerical analysis, investigating coupling combinations,
parameter identifiability, and the robustness of the results
under different physical assumptions.
Finally, Section~\ref{sec:conclusion} summarizes
our main findings and conclusions.

\section{Physical Model}
\label{sec:response}
We assume mediator mass $m_X=16.88\,\MeV$ \cite{AriasAragon2026}.
We use $\hbar=c=1$, where $\hbar$ is the reduced Planck constant and $c$ is the speed of light; numerically, $\hbar c=197.326980459\,\MeV\,\fm$. The vector interaction is
\begin{equation}
 \mathcal L_X=eX_\lambda\sum_{f=e,\mu,p,n}\epsilon_f\bar f\gamma^\lambda f.
 \label{eq:lag}
\end{equation}
Here $\mathcal L_X$ is the Lagrangian density, $X_\lambda$ the mediator field, and $f$ a fermion field with Dirac adjoint $\bar f=f^\dagger\gamma^0$. The $\gamma^\lambda$ are Dirac matrices; $\lambda$ is a contracted Lorentz index. The elementary charge $e>0$ defines the fine-structure constant $\alpha=e^2/(4\pi)$. Real dimensionless couplings $\epsilon_f$ have independent signs; $e,\mu,p,n$ label electrons, muons, protons and neutrons.

For lepton $\ell=e,\mu$ and nucleus $A$ with proton and neutron numbers $Z,N$, the momentum-space potential is
\begin{equation}
 \widetilde V_X(\bm q)=\frac{4\pi\alpha\epsilon_\ell[Z\epsp F_p(q)+N\epsn F_n(q)]}{q^2+\mx^2},
 \qquad \epsilon_A=Z\epsp+N\epsn.
 \label{eq:potential}
\end{equation}
Here $q=|\bm q|$ is the transferred momentum magnitude, $F_p,F_n$ are proton and neutron current form factors normalized to unity at $q=0$, and $\epsilon_A$ is the nuclear vector charge. Point sources give $V_X(r)=\alpha\epsilon_\ell\epsilon_A e^{-\mx r}/r$ at lepton--nucleus separation $r$. Equal-sign charges repel. Factoring out $\epsilon_A$ requires $F_p=F_n$; otherwise $\epsilon_A=0$ need not cancel the potential. Electromagnetic-like profiles are assumptions.

Let $m_\ell,M_A$ be the lepton and nuclear masses. Define the reduced mass $\mu_{\ell A}=m_\ell M_A/(m_\ell+M_A)$ and Coulomb length $a_{\ell A}=(Z\alpha\mu_{\ell A})^{-1}$. The point-source muonic response and our reference dipole form factor are
\begin{equation}
 K_A^{\rm point}=\frac{\alpha a_{\mu A}\mx^2}{2(1+a_{\mu A}\mx)^4},
 \qquad F_{\rm dip}(q)=(1+q^2R_A^2/12)^{-2}.
 \label{eq:point}
\end{equation}
Here $K_A$ is the positive $2S-2P$ shift per unit $\epsm\epsilon_A$; $S,P$ denote orbital angular momenta zero and one. The fixed source radius $R_A$ is its root-mean-square radius. Dipole sources and point-Coulomb wave functions give Table~\ref{tab:atomic}.

\begin{table}[!ht]
\centering
\caption{Reference source radii, SM field coefficients \cite{Pachucki2024}, and dipole muonic responses. H and D denote hydrogen and deuterium.}
\label{tab:atomic}
\small
\begin{tabular}{@{}lrrrr@{}}
\toprule
Nucleus $A$ & H & D & ${}^3$He & ${}^4$He\\
\midrule
$R_A$ [fm] & 0.84060 & 2.12758 & 1.97007 & 1.6786\\
$C_A$ [$\mev\,\fm^{-2}$] & $-5.2259$ & $-6.1074$ & $-103.383$ & $-106.209$\\
$K_A$ [meV] & 3627.551 & 4199.960 & 29937.847 & 30679.725\\
\bottomrule
\end{tabular}
\end{table}

The conventional Lamb shift $L_A=E_{2P}-E_{2S}$, with level energies $E_{2P},E_{2S}$, obeys
\begin{equation}
 L_A^{\rm obs}=L_A^0+C_A r_{A,\rm true}^2-K_A\epsm\epsilon_A,
 \qquad r_{A,\rm SM}^2=r_{A,\rm true}^2+b_A\epsm\epsilon_A,
 \quad b_A=K_A/|C_A|.
 \label{eq:bias}
\end{equation}
Here $L_A^0$ contains radius-independent SM terms, $C_A<0$ is the field coefficient, and $r_{A,\rm true}$ and $r_{A,\rm SM}$ are physical and SM-extracted charge radii.

For electronic H and D, write the new-force $1S-2S$ frequency shift as $\delta\nu_A=-K_{21,A}\epse\epsilon_A/h$, with Planck constant $h=2\pi\hbar$. The dipole responses are $K_{21,H}=0.00463608447\,\mev$ and $K_{21,D}=0.00463985102\,\mev$. The isotope convention $\nu_D-\nu_H$ gives $C_{HD}=-1369.5\,\mathrm{kHz}\,\fm^{-2}$ \cite{Pachucki2018,Potvliege2025}. Define $d_{HD}=(r_D^2-r_p^2)_{\mu,\rm SM}-(r_D^2-r_p^2)_{e,\rm SM}$, where $r_p,r_D$ are proton and deuteron charge radii. Define $d_{He}$ analogously using $r_3^2-r_4^2$, with $3,4$ denoting helium isotopes. Their predicted new-force contributions are
\begin{align}
 P_{HD}&=\epsm(A_\mu\epsp+B_\mu\epsn)-\epse(A_e\epsp+B_e\epsn),\label{eq:hd}\\
 P_{He}&=\epsm(A_{\mu,He}\epsp+B_{\mu,He}\epsn)+B_{e,He}\epse\epsn.\label{eq:he}
\end{align}
The response coefficients, all in $\fm^2$, are
\begin{align}
 (A_\mu,B_\mu)&=(b_D-b_H,b_D)=(-6.4649,687.6838),\nonumber\\
 (A_e,B_e)&=(K_{21,D}-K_{21,H},K_{21,D})/(h|C_{HD}|)=(0.6650,819.2121),\label{eq:coefficients}\\
 (A_{\mu,He},B_{\mu,He},B_{e,He})&=(2(b_3-b_4),b_3-2b_4,3(\hbar c/\mx)^2)=(1.4402,-288.1417,409.9674).\nonumber
\end{align}
Equation~\eqref{eq:coefficients} uses consistent energy--frequency units. For electronic helium, the contact approximation gives $\delta E_X/(\partial E/\partial r_A^2)=6\epse\epsilon_A/(Z\mx^2)$, where $E$ is the transition energy and $\delta E_X$ its new-force shift. Equation~\eqref{eq:he} assumes isotope-independent electronic densities, neglecting finite-range, recoil and higher field-shift corrections. The electron--electron term cancels at infinite nuclear mass; its recoil remainder is omitted.

Define $\bm n=(\epsp,\epsn)^T$ and $\bm P=(P_{HD},P_{He})^T$, where $T$ denotes transposition. The response matrices are
\begin{equation}
 U=\begin{pmatrix}A_\mu&B_\mu\\A_{\mu,He}&B_{\mu,He}\end{pmatrix},\qquad
 E=\begin{pmatrix}A_e&B_e\\0&-B_{e,He}\end{pmatrix},\qquad
 \bm P=(\epsm U-\epse E)\bm n.
 \label{eq:matrices}
\end{equation}
Rows label H--D and helium; columns label protons and neutrons. Entries have units $\fm^2$. For fixed one-body responses, spectroscopy is invariant under $(\epse,\epsm,\epsp,\epsn)\mapsto(\epse/s,\epsm/s,s\epsp,s\epsn)$ for any dimensionless scale $s>0$.

For lepton $\ell$, define the magnetic anomaly $a_\ell=(g_\ell-2)/2$, where $g_\ell$ is the gyromagnetic factor. The one-loop vector contribution is \cite{Leveille1978,Pospelov2009}
\begin{equation}
 \Delta a_\ell^X=C_\ell\epsilon_\ell^2,\qquad
 C_\ell=\frac{\alpha}{2\pi}\int_0^1\dd z\,
 \frac{2z(1-z)^2}{(1-z)^2+z(\mx/m_\ell)^2},\qquad
 C_e=6.999208\times10^{-7},\quad C_\mu=7.899579\times10^{-4}.
 \label{eq:g2}
\end{equation}
Here $z$ is dimensionless and $C_\ell>0$ is the loop coefficient. Magnetic moments leave coupling signs undetermined. This leading-order treatment assumes small shifts, perturbative lepton interactions and controlled nuclear sources. The usual condition $\alpha\epsilon_f^2\ll1$ is not by itself sufficient for a fixed-nucleus treatment: a mediator exchange between nucleons can change the nuclear wave function even when lepton loops remain weak. Nuclear polarization induced by $X$, relativistic, radiative and mixed photon--vector corrections are omitted. We quantify a necessary consistency diagnostic in Sec.~\ref{sec:physical_domain}.

\section{Numerical Analysis}
\label{sec:identifiability}

The numerical study proceeds in four steps. We first construct a conditional likelihood from published electronic and muonic isotope-radius differences and lepton magnetic moments. We then establish which coupling products are tested by the isotope comparisons, including a direction in which the hydrogen--deuterium neutron response cancels. Next, we distinguish the local sensitivity of all four observables from the global behaviour of the likelihood when the nucleon couplings are unrestricted. Finally, we repeat the profiling with explicit, illustrative restrictions on the additional nucleon--nucleon potential. Keeping these steps separate is essential: a preference for a coupling product, an invertible local response matrix, and a physically meaningful individual-coupling bound are different statements.

\subsection{Experimental inputs and conditional likelihood}
\label{sec:inputs}
We adopt Gaussian errors and fixed SM inputs. Muonic H and D radii \cite{Antognini2013,Pohl2016} and the electronic isotope shift \cite{Parthey2010} give squared-radius differences $3.8200(31)$ and $3.8207(3)\,\fm^2$ \cite{Pachucki2024,Pachucki2018}. Parentheses denote one-standard-deviation errors on the last digits. Hence $d_{HD}=(-0.0007\pm0.003114)\,\fm^2$, including correlated nuclear corrections. The muonic helium difference \cite{Krauth2021,Schuhmann2025} is $1.0626(29)\,\fm^2$, with correlated nuclear errors \cite{LiMuli2025}. Electronic alternatives (Table~\ref{tab:helium}) are analyzed separately: Cancio Pastor with updated isotope theory \cite{Cancio2012,PachuckiHe2017,Potvliege2025}; singlet--triplet with revised recoil terms \cite{Steinebach2026,PachuckiRecoil2025} replacing the earlier smaller error \cite{PachuckiHFS2024}; Wen with corrected helium-4 postselection and the same helium-3 input as Cancio Pastor \cite{Wen2025}.

\begin{table}[!ht]
\centering
\caption{Electronic helium radius differences $\delta r^2_{e,He}=(r_3^2-r_4^2)_{e,\rm SM}$ and H--D compatibility at $\epse=\epsp=0$. Tension is $\sqrt{\chi^2_{\min}}$ for independent inputs.}
\label{tab:helium}
\small
\begin{tabular}{@{}lrrr@{}}
\toprule
Electronic input & $\delta r^2_{e,He}$ [fm$^2$] & $\chi^2_{\min}$ & Tension\\
\midrule
Steinebach \cite{Steinebach2026} & $1.0676(10)$ & 2.521 & 1.59\\
Cancio Pastor \cite{Cancio2012} & $1.0690(30)$ & 2.344 & 1.53\\
Wen \cite{Wen2025} & $1.0733(21)$ & 8.321 & 2.88\\
\bottomrule
\end{tabular}
\end{table}

Table~\ref{tab:helium} illustrates why the choice of electronic helium input must be stated rather than averaged away. The three determinations enter the same observable, $d_{He}=(r_3^2-r_4^2)_{\mu,\rm SM}-(r_3^2-r_4^2)_{e,\rm SM}$, but lead to different residual central values and uncertainties. In the table, the displayed $\chi^2_{\min}$ is obtained only on the restricted slice $\epse=\epsp=0$, where both isotope comparisons test a single neutron-coupling product; it does \emph{not} describe the minimum of the unrestricted four-coupling model. This distinction will be central below.

The data vector $\bm d=(d_{HD},d_{He})^T$ has reference covariance $V_0=\operatorname{diag}(\sigma_{HD}^2,\sigma_{He}^2)$, where $\sigma_{HD},\sigma_{He}$ combine published errors in quadrature. We set $V=V_0$ for the baseline statistic $\chi^2_{\rm spec}=(\bm P-\bm d)^TV^{-1}(\bm P-\bm d)$. This is a \emph{conditional} four-input compressed likelihood, not a global refit of the underlying transition frequencies: the two isotope residuals share theory ingredients and may have an unreported cross-covariance. We do not assign an empirical value to that covariance.

Define residual $y_\ell=a_\ell^{\rm exp}-a_\ell^{\rm SM}$ and error $\sigma_\ell$. Muon measurements and the lattice-based SM prediction give $y_\mu=(38.5\pm63.673)\times10^{-11}$ \cite{Aguillard2025,Aguillard2026,Aliberti2025}. This $0.605$-standard-deviation residual does not establish a nonzero coupling. Earlier dispersive predictions give different residuals.

The electron measurement \cite{Fan2023} is combined separately with cesium (Cs) and rubidium (Rb) recoil values of $\alpha$ \cite{Parker2018,Morel2020}. SM coefficients \cite{Mohr2025} and Volkov's five-loop result \cite{Volkov2024} give $y_e^{\rm Cs}=(-95.92\pm26.29)\times10^{-14}$ and $y_e^{\rm Rb}=(39.27\pm16.03)\times10^{-14}$. The alternative calculation \cite{Aoyama2025} changes either by $0.145\times10^{-14}$. A positive vector contribution cannot explain Cs; Rb favors it at $2.45$ standard deviations. Recoil-derived $\alpha$ remains external. We use
\begin{equation}
 \chi^2=\chi^2_{\rm spec}+\chi^2_{g-2},\qquad
 \chi^2_{g-2}=\sum_{\ell=e,\mu}(C_\ell\epsilon_\ell^2-y_\ell)^2/\sigma_\ell^2.
 \label{eq:joint}
\end{equation}
The first term compares two measured isotope residuals with the two predictions $P_{HD}$ and $P_{He}$ from Eqs.~\eqref{eq:hd} and~\eqref{eq:he}. The second term fixes the preferred \emph{magnitudes} of the lepton couplings through $C_\ell\epsilon_\ell^2$ without selecting their signs. It therefore supplies information unavailable to spectroscopy alone, but does not directly measure either nucleon coupling. A small total $\chi^2$ is agreement within this specified compressed model, not evidence for $X_{17}$ in isolation.

Experimental and SM errors add in quadrature, and Cs and Rb are always treated as alternatives. Because signs, boundaries, singular parameter points and theory uncertainties violate the regular assumptions behind simple asymptotic likelihood-ratio tests, we label the reported $\Delta\chi^2$ sublevel sets as mathematical profiles rather than nominally calibrated confidence regions.

Reference inputs also affect the summaries. Write the isotope frequency $\nu_{\rm iso}=\nu_0(\bm\eta)+C\delta r_{\rm true}^2+DS_{\rm true}+\delta\nu_X$. Here $\nu_0$ is the radius-independent prediction with SM inputs $\bm\eta$; $C,D$ are field coefficients and $\delta\nu_X$ the new-force shift. For isotope charge radii $r_1,r_2$, define $\delta r^2=r_1^2-r_2^2$ and $S=r_1^2+r_2^2$. The measured residual $d$ and predicted response $P$ satisfy
\begin{equation}
 d=P+\mathcal R,\qquad
 \mathcal R=\frac{\nu_0(\bm\eta_{\rm ref})-\nu_0(\bm\eta_{\rm true})}{C}
 +\frac{D}{C}(S_{\rm ref}-S_{\rm true}).
 \label{eq:nuisance}
\end{equation}
Here $\mathcal R$ is the reference-input correction; ``ref'' and ``true'' label adopted and physical values. The baseline sets $\mathcal R=0$ and therefore conditions on the reference inputs. Physically, the same measured transition frequency can produce a different reported radius difference if the reference Standard-Model constants or the charge-radius sum are changed. The compressed residual $d$ is thus not an independent, theory-free observable. Were a justified Gaussian nuisance covariance $S_{\mathcal R}$ available, analytical marginalization of the additive nuisance would replace $V_0$ by $V_0+S_{\mathcal R}$ (with a covariance normalization when the covariance is parameter-dependent). Such a covariance cannot be inferred from the summary measurements supplied here, so we do not manufacture one. Instead we test illustrative residual correlations $\rho=\pm0.5$ and keep the helium experiments separate. Helium singlet--triplet coefficients are $(C,D)=(-214.758,0.017)\,\mathrm{kHz}\,\fm^{-2}$ \cite{Steinebach2026,PachuckiRecoil2025}. A muonic reference sum shifts helium muonic coefficients by $(-0.0916,-0.0687)\,\fm^2$; unequal electronic field coefficients add $(-0.1298,-0.0974)\,\fm^2$ to the helium row of $E$. These modifications illustrate a reference dependence and are not fitted nuisance shifts.

\subsection{Isotope compatibility, neutron cancellation and source sensitivity}
\label{sec:isotope}

A useful first diagnostic is to switch off the electronic and proton couplings, leaving the single product $x=\epsm\epsn$. This is a deliberately restricted test of whether the two isotope measurements favour a consistent response, not an assumption imposed in the full fit. On $\epse=\epsp=0$, both spectra measure $x=\epsm\epsn$. For $i=HD,He$, define $x_i=d_i/B_i$ and $s_i=\sigma_i/|B_i|$, with $B_{HD}=B_\mu$, $B_{He}=B_{\mu,He}$. The minimum $\chi^2_{\min}=(x_{HD}-x_{He})^2/(s_{HD}^2+s_{He}^2)$ has one degree of freedom, provided the two input errors are independent. Hydrogen--deuterium gives $x_{HD}=(-1.018\pm4.529)\times10^{-6}$. With the Steinebach input, the conditional compatibility tension is $1.588$ standard deviations; the two-sided Wen alternative gives $2.885$ standard deviations (probability $0.00392$) and joint product $x=(3.46\pm4.26)\times10^{-6}$, consistent with zero. A residual correlation $\rho$ contributes $+2\rho s_{HD}s_{He}$ to the variance of the difference because $B_{HD}B_{He}<0$. Assumed $\rho=(-0.5,0,+0.5)$ changes the Steinebach tension to $(1.99,1.59,1.36)$ standard deviations and the Wen tension to $(3.50,2.88,2.51)$. These are sensitivity scenarios, not measured correlations. Consequently, the larger Wen tension in Table~\ref{tab:helium} should be read as input-dependent incompatibility within this one-product slice, not as a model-independent anomaly or a detection of the mediator. Once both lepton and both nucleon couplings vary, the two isotope central values can generally be fitted simultaneously; the question then becomes whether the required couplings are uniquely and physically determined.

\begin{figure}[t]
\centering
\includegraphics[width=\textwidth]{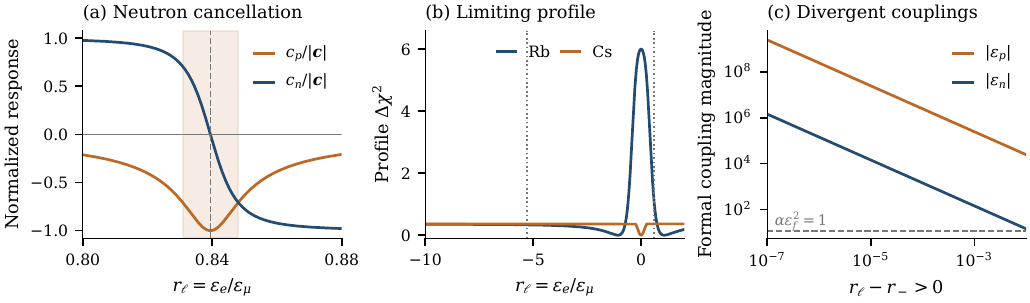}
\caption{Spectroscopic sensitivity and the formally unrestricted profile. (a) Normalized H--D proton and neutron coefficients as functions of $r_\ell=\epse/\epsm$; shading marks $|c_p|>|c_n|$, and the dashed vertical line marks $c_n=0$. The plotted coefficients do not include the unknown nucleon couplings. (b) Profile $\Delta\chi^2$ after exact fitting of both isotope central values away from singular ratios and minimizing over the lepton magnitude, for alternative Rb/Cs inputs; dotted lines mark $\det W=0$. Limits approached at these lines need not equal values at the singular points. (c) Divergence of $|\epsp|$ and $|\epsn|$ for the Rb--Wen exact-spectra sequence approaching $r_-$ from above. The horizontal line $\alpha\epsilon_f^2=1$ is a perturbative warning, not an experimental limit.}
\label{fig:identifiability}
\end{figure}

For $\epsm\ne0$, define the lepton ratio $r_\ell=\epse/\epsm$. Then $P_{HD}=\epsm(c_p\epsp+c_n\epsn)$, with proton and neutron coefficients
\begin{equation}
 c_p=A_\mu-r_\ell A_e=(-6.4649-0.6650r_\ell)\,\fm^2,\qquad
 c_n=B_\mu-r_\ell B_e=(687.6838-819.2121r_\ell)\,\fm^2.
 \label{eq:direction}
\end{equation}
At $r_\ell=0.839445$, $c_n=0$ while $c_p=-7.0231\,\fm^2$; $0.830879<r_\ell<0.848025$ gives $|c_p|>|c_n|$ (Fig.~\ref{fig:identifiability}a). Physical neutron dominance requires $|c_n\epsn|>|c_p\epsp|$. Relative to the independent-lepton minimum, cancellation costs $\Delta\chi^2_{g-2}=0.1264$ for Rb at $|\epsm|=8.6781\times10^{-4}$, and $0.3656$ for Cs in the zero-lepton limit. Thus hydrogen--deuterium supplies no universal neutron bound. Equation~\eqref{eq:hd} remains regular at $\epsm=0$.

Figure~\ref{fig:identifiability}(a) makes the cancellation transparent by plotting $c_p/|\bm c|$ and $c_n/|\bm c|$ against the signed lepton-coupling ratio, where $|\bm c|=(c_p^2+c_n^2)^{1/2}$. The neutron coefficient crosses zero at the dashed line, whereas the proton coefficient remains finite. The shaded interval denotes the region where the \emph{coefficient} of the proton coupling has a larger absolute magnitude than that of the neutron coupling. It does not establish that the proton term dominates the measured isotope shift, because the actual contributions also contain $\epsp$ and $\epsn$. The physical consequence is narrower, but important: even highly accurate hydrogen--deuterium spectroscopy cannot supply a ratio-independent constraint on $\epsn$ when the electronic and muonic responses are allowed to cancel.

Table~\ref{tab:sensitivity} compares sources, including the Gaussian form factor $F_{\rm G}(q)=\exp(-q^2R_A^2/6)$. Extended electromagnetic (EM) wave functions reduce muonic responses by $0.018$--$0.24\%$, but change the small H--D and helium proton differences by $10.5\%$ and $24.7\%$. Independent $\pm10\%$ source-radius variations give $A_\mu\in[-6.616,-6.323]$ and $A_{\mu,He}\in[1.094,1.782]\,\fm^2$. These test sensitivity, not vector-current uncertainty. The relatively small changes in the large neutron coefficients $B_\mu$ and $B_{\mu,He}$ should not obscure the more substantial fractional changes in the proton combinations $A_\mu$ and $A_{\mu,He}$: the latter are differences of larger atomic responses and are therefore particularly susceptible to changes in the source prescription. This matters for the location and conditioning of the cancellations, rather than establishing a new statistical uncertainty distribution. Coordinate- and momentum-space evaluations agree within $2\times10^{-8}$ relatively, checking integration accuracy only.

\begin{table}[t]
\centering
\caption{Muonic response coefficients in $\fm^2$. The first three rows use point-Coulomb wave functions; the last uses an extended dipole electromagnetic (EM) potential. Electronic coefficients remain fixed.}
\label{tab:sensitivity}
\small
\begin{tabular}{@{}lrrrr@{}}
\toprule
Prescription & $A_\mu$ & $B_\mu$ & $A_{\mu,He}$ & $B_{\mu,He}$\\
\midrule
Point source & $-5.9133$ & 688.3368 & 1.7077 & $-288.3866$\\
Dipole source & $-6.4649$ & 687.6838 & 1.4402 & $-288.1417$\\
Gaussian source & $-6.4711$ & 687.6771 & 1.4363 & $-288.1403$\\
Extended EM & $-7.1424$ & 686.8794 & 1.0842 & $-287.8011$\\
\bottomrule
\end{tabular}
\end{table}

\subsection{Identifiability and formal likelihood directions}
\label{sec:global_profiles}

The cancellation of a particular isotope coefficient is not the same as a failure to determine all four couplings locally. To test the latter, consider the full map from the four signed couplings to the two isotope residuals and the two magnetic-moment shifts. Let $\bm f=(P_{HD},P_{He},C_e\epse^2,C_\mu\epsm^2)$ be the observable vector and $\bm\epsilon=(\epse,\epsm,\epsp,\epsn)$ the parameter vector. The Jacobian $J_{ij}=\partial f_i/\partial\epsilon_j$, with component indices $i,j=1,\ldots,4$, satisfies
\begin{equation}
 \det J=4C_eC_\mu\epse\epsm\det(\epsm U-\epse E).
 \label{eq:jacobian}
\end{equation}
For nonzero lepton couplings and spectral determinant, the rank is four. H--D plus both moments has rank three; spectroscopy alone has rank two. Helium restores generic local identifiability. Zero-lepton points are nonregular; sign ambiguities persist. Full rank here means that sufficiently small parameter displacements generically produce distinguishable changes in the four predictions. It does \emph{not} bound the parameter space far from the expansion point, and it does not eliminate the discrete sign branches of the lepton couplings. This is why the rank test must be supplemented by a global profile analysis.

Define $W(r_\ell)=U-r_\ell E$. Away from its singularities and from $\epsm=0$, the nucleon solution that fits both spectral central values is
\begin{equation}
 \widehat{\bm n}=\epsm^{-1}W(r_\ell)^{-1}\bm d,\qquad
 \det W(r_\ell)/\fm^4=872.4086-1278.9509r_\ell-272.6372r_\ell^2.
 \label{eq:nprofile}
\end{equation}
For nonsingular $V$, profiling (minimizing) over nucleon couplings removes spectral residuals. Conditional tensions do not exclude the unrestricted model. At determinant roots $r_+=0.60429$ and $r_-=-5.29532$, spectral directions align, but every adopted data vector lies outside the column space. Approaching either root gives divergent $\widehat{\bm n}$ and exact fits. At the root a residual remains, so the limiting infimum differs from the value there. Geometrically, the proton and neutron response columns become linearly dependent at each root. On approaching a root, increasingly large proton and neutron contributions can nearly cancel while leaving the two predicted isotope shifts fixed. Thus the divergence is a property of the inverse response map, not an indication that spectroscopy favours a physically large nucleon coupling.

Both roots remain simple and real with extended EM wave functions ($0.647,-7.43$) or helium reference corrections ($0.606,-9.28$). As an additional purely algebraic stress test, independently changing the four entries of the muonic-response matrix $U$ by $\pm5\%$ in all 16 sign combinations preserves two real roots, with the negative root ranging from $-6.06$ to $-4.50$ and the positive root from $0.480$ to $0.702$. This establishes stability against those \emph{response-coefficient perturbations}; it is not a calculation of separate proton and neutron form factors, their correlations or nuclear structure. Varying $\mx$ over $16.83$--$16.93\,\MeV$ changes dominant coefficients by about $0.6\%$. Full nuclear-current robustness is not established.

At fixed $r_\ell$, minimize over the squared muon coupling $u=\epsm^2\geq0$:
\begin{equation}
 \widehat u(r_\ell)=\max\!\left[0,
 \frac{C_\mu y_\mu/\sigma_\mu^2+C_e r_\ell^2y_e/\sigma_e^2}
 {C_\mu^2/\sigma_\mu^2+C_e^2r_\ell^4/\sigma_e^2}\right].
 \label{eq:qprofile}
\end{equation}
For Rb, root limits lie $\Delta\chi^2=0.3362$ at $r_-$ and $1.0555$ at $r_+$ above the global infimum. The formal set $\Delta\chi^2\leq1$ is unbounded (Fig.~\ref{fig:identifiability}b,c), without an assigned confidence level. For Cs, vanishing leptons at fixed nonsingular ratio preserve exact fits under coupling rescaling, at limiting cost $0.3656$. Rb instead diverges through response alignment; its zero-lepton penalty is about $6.37$. Exact nonzero lepton magnitudes away from the roots would fix nucleon couplings on each sign branch.

Figures~\ref{fig:identifiability}(b) and~\ref{fig:identifiability}(c) display two aspects of this global result. In panel (b), the ratio-dependent profile is obtained by fitting \emph{both} isotope central values exactly whenever $W(r_\ell)$ is invertible and then minimizing the magnetic-moment contribution over the lepton magnitude. The horizontal axis is the signed ratio $r_\ell$, not the magnitude of an individual coupling, while the vertical axis is the increase in $\chi^2$ relative to the unrestricted minimum. The Rb and Cs curves differ because a positive vector contribution can accommodate the Rb electron residual but not the negative Cs residual. The dotted vertical lines mark the singular ratios of $W$; the plotted values approached near them are limiting profile values and need not equal the value at exact singularity.

Panel (c) isolates the approach to the negative root, plotting the absolute proton and neutron couplings of an Rb fit to the H--D and Wen helium central values against the positive distance $r_\ell-r_-$. Both couplings increase without bound as this distance tends to zero, even though the isotope residuals remain exactly fitted along the sequence. The dashed horizontal reference corresponds to $\alpha\epsilon_f^2=1$, indicating where the assumed perturbative expansion is plainly unreliable; the fixed nuclear-source description can fail even earlier. Panels (b) and (c) therefore establish \emph{formal} unboundedness of the compressed profile, not a physically permitted infinite-coupling region or a calibrated confidence interval.

The divergent sequences eventually violate the controlled-source and perturbative assumptions of Sec.~\ref{sec:response}. A finite-coupling domain can regularize these mathematical tails, but imposing a numerical ceiling without stating its physical origin would turn a model assumption into a spurious experimental exclusion. We therefore make the assumed ceiling explicit in the following diagnostic. Complementary probes include hyperfine splittings \cite{Jentschura2020,Lin2026}, pion and kaon decays \cite{Hostert2023,Hostert2026}, and NA64, KLOE and NA48/2 searches \cite{NA642020,KLOE2015,NA4822015}. Translating production and decay searches into this four-coupling space requires additional quark and neutrino couplings, branching ratios and lifetimes; none are silently folded into Eq.~\eqref{eq:joint}.

\subsection{Finite-coupling stress tests and physical interpretation}
\label{sec:physical_domain}
A weak lepton interaction need not imply a negligible change of nuclear binding. To make this point quantitative without pretending to perform a nuclear many-body fit, consider the tree-level Yukawa potential of the same vector mediator between two nucleons $i,j\in\{p,n\}$ at a reference separation $r_0$:
\begin{equation}
 |V_{ij}(r_0)|=\alpha|\epsilon_i\epsilon_j|\frac{\hbar c}{r_0}
 \exp\!\left(-\frac{m_Xr_0}{\hbar c}\right),\qquad
 |V_{ij}(1\,\fm)|=1.32191|\epsilon_i\epsilon_j|\,\MeV.
 \label{eq:NNdiagnostic}
\end{equation}
The source profile and short-distance nuclear dynamics have been suppressed in this deliberately simple diagnostic. Imposing the illustrative requirement $\max_{i,j}|V_{ij}(r_0)|\leq\Lambda_N$ at $r_0=1\,\fm$ is equivalent, for these point potentials, to $\max(|\epsp|,|\epsn|)\leq[\Lambda_N/(1.32191\,\MeV)]^{1/2}$. We use $\Lambda_N=0.1$ and $1\,\MeV$ as \emph{stress thresholds}, not as nuclear-theory errors, experimental exclusions or confidence limits. Controlling this one-distance proxy is neither necessary nor sufficient for a reliable many-body calculation; the physical potential changes nuclear wave functions and finite nuclear currents, effects omitted from the atomic response.

For illustration, the Rb central value of $a_e$, central $a_\mu$ and the Steinebach and H--D radius central values admit an exact fit of the compressed one-body equations with $(\epse,\epsm)=(7.49,6.98)\times10^{-4}$ and $(\epsp,\epsn)=(1.87,-0.0649)$. In this \emph{unrestricted} solution, the point-proton potential has magnitude $|V_{pp}(1\,\fm)|=4.63\,\MeV$. This is a warning about the fixed-source approximation, not proof that the solution is excluded by nuclear data.

To study the sensitivity to finite nuclear couplings, we minimize Eq.~\eqref{eq:joint} with $|\epsp|,|\epsn|\leq\epsilon_N^{\max}$, solving the two-dimensional bounded linear spectral least-squares problem for every lepton-coupling pair, then minimizing the remaining nonnegative objective. The reference results, for zero residual cross-covariance, are in Table~\ref{tab:restricted}. Each entry gives the minimized total $\chi^2$; the unrestricted Rb fit has infimum zero for each helium selection. An assumed $\Lambda_N=0.1\,\MeV$ still permits a Steinebach fit at $\chi^2\simeq0.107$, with $\epsp\simeq0.275$ and $\epsn\simeq-0.100$; thus excluding the large-coupling \emph{central-value solution} does not exclude good alternate fits. The change in minimum is a numerical likelihood cost, not a calibrated statistical significance. Table~\ref{tab:restricted} also shows that relaxing the stress threshold from $0.1$ to $1\,\MeV$ decreases the minimum $\chi^2$ for both helium alternatives, as expected when the allowed parameter domain becomes larger. Small conditional fit penalties should not, however, be confused with a proof that the corresponding nuclear wave functions and source responses remain unchanged.

\begin{table}[!ht]
\centering
\caption{Conditional Rb minima under point-potential diagnostics at $r_0=1\,\fm$. Thresholds are illustrative assumptions, not nuclear limits or confidence bounds. Electronic helium inputs are alternatives, never combined.}
\label{tab:restricted}
\small
\begin{tabular}{@{}lrrr@{}}
\toprule
$\Lambda_N$ [MeV] & $\epsilon_N^{\max}$ & $\min\chi^2$ (Steinebach) & $\min\chi^2$ (Wen)\\
\midrule
0.1 & 0.275 & 0.107 & 0.117\\
1.0 & 0.870 & 0.045 & 0.085\\
Unrestricted & --- & 0 (infimum) & 0 (infimum)\\
\bottomrule
\end{tabular}
\end{table}

\begin{figure}[!ht]
\centering
\includegraphics[width=\textwidth]{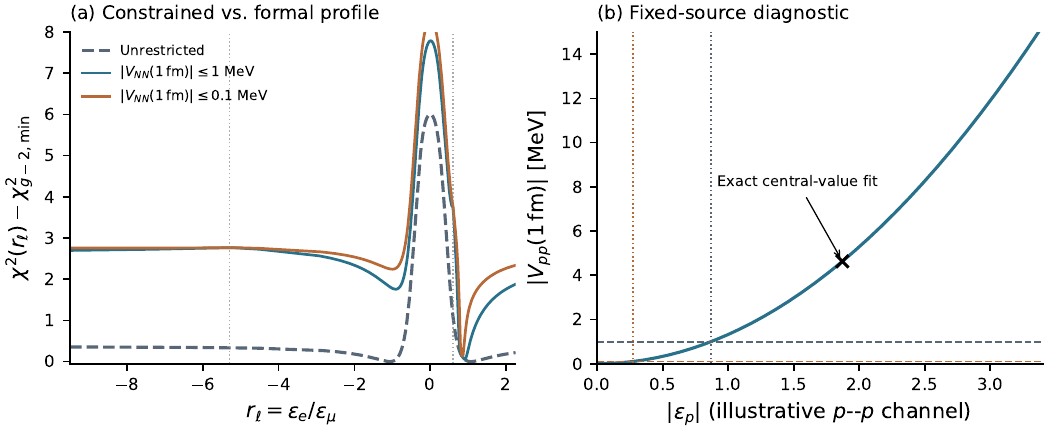}
\caption{Finite-coupling diagnostics for the Rb electron input and Steinebach helium input. (a) Total $\chi^2$ at fixed signed lepton ratio, minimized over the remaining couplings; the dashed curve is unrestricted and the solid curves impose $\max_{i,j}|V_{ij}(1\,\fm)|\leq0.1$ or $1\,\MeV$ in the point-potential proxy. All curves subtract the unrestricted $g-2$ minimum, rather than a calibrated confidence threshold. Dotted vertical lines mark $\det W=0$. (b) $p$--$p$ Yukawa-potential magnitude at $1\,\fm$ versus $|\epsp|$; horizontal dashed levels and vertical dotted lines indicate the assumed $0.1$ and $1\,\MeV$ diagnostic cutoffs. The cross marks the unrestricted exact-central-value fit. Neither threshold represents an empirical nuclear exclusion.}
\label{fig:physical_domain}
\end{figure}

Figure~\ref{fig:physical_domain}(a) compares the \emph{ratio-dependent} profile in the unrestricted model (dashed curve) with profiles obtained after imposing the two illustrative finite-coupling ceilings. For each chosen $r_\ell$, the muon-coupling magnitude and the bounded proton and neutron couplings are re-optimized. The vertical quantity is the total $\chi^2$ minus the unconstrained minimum of the $g-2$ contribution, so these curves are neither likelihood-ratio confidence contours nor directly comparable to a universal $\Delta\chi^2$ critical value. Near response-alignment regions, finite couplings can no longer compensate freely for nearly dependent isotope responses, and the restricted curves rise relative to the formal profile. Elsewhere, good fits remain possible despite the imposed ceilings. The dotted vertical lines indicate $r_\pm$, not experimental exclusions.

Figure~\ref{fig:physical_domain}(b) shows the origin and numerical scale of these ceilings for the illustrative proton--proton channel. The solid curve is the quadratic relation $|V_{pp}(1\,\fm)|=1.32191\,\MeV\,|\epsp|^2$, while the horizontal dashed lines at $0.1$ and $1\,\MeV$ and their corresponding vertical dotted lines identify the chosen proxy cutoffs $|\epsp|=0.275$ and $0.870$. The cross marks the unrestricted exact-central-value fit, $|\epsp|\simeq1.87$, for which $|V_{pp}(1\,\fm)|\simeq4.63\,\MeV$. Its location illustrates why an excellent atomic fit need not be consistent with an unchanged nuclear source. The full restricted optimization bounds \emph{both} $\epsp$ and $\epsn$, whereas this panel depicts only the $p$--$p$ contribution; no microscopic nuclear exclusion is inferred from the plot.

The cancellation of H--D neutron sensitivity and the generic Jacobian rank are properties of the specified response coefficients. Their numerical cancellation ratios move if proton and neutron form factors change separately; neither the $\pm5\%$ response test nor the finite-coupling exercise supplies a self-consistent calculation of $F_p(q)$ and $F_n(q)$. We therefore do not quote model-independent individual limits on $\epsp$ or $\epsn$. Such limits require a nuclear response calculation, justified priors or supplementary observables and, for frequentist confidence statements, coverage studies. Gaussian profiling alone is insufficient near the nonregular lepton-zero boundaries and singular ratios.

\section{Conclusions}
\label{sec:conclusion}

Precision atomic spectroscopy represents one of the most sensitive approaches to investigating light mediators and possible deviations from the Standard Model. However, the interpretation of increasingly precise measurements requires a careful distinction between the quantities directly probed by experiments and the fundamental parameters one aims to constrain. In this work, we have investigated this distinction for the hypothetical $X_{17}$ vector boson, considering four independent electron, muon, proton, and neutron couplings. By combining electronic and muonic spectroscopy of hydrogen, deuterium, and helium with the anomalous magnetic moments of charged leptons, we have established the structure of the
coupling combinations accessible to these complementary observables and identified the conditions under which individual parameters can be inferred.

A first important result concerns the sensitivity of isotope comparisons to the neutron coupling. Although hydrogen--deuterium spectroscopy is particularly attractive for probing interactionswith neutrons, we have demonstrated that its sensitivity is not universal. For specific relations between electron and muon couplings, the corresponding neutron contributions can cancel, leaving the observable predominantly sensitive to the proton sector. This effect does not require the neutron interaction itself to vanish, but follows from the interplay between electronic and muonic responses. The inclusion of helium isotope measurements introduces a complementary nuclear response, while the electron and muon anomalous magnetic moments provide independent information on the leptonic coupling magnitudes. We have shown that their combination generically restores local identifiability of the four couplings, demonstrating the importance of combining observables with different physical sensitivities.

The most significant result of our analysis is that this local identifiability does not necessarily translate into finite constraints on individual couplings. Despite the generic full rank of the combined sensitivity matrix, the unrestricted statistical model admits parameter directions along which nucleon couplings become arbitrarily large without a corresponding deterioration sufficient to exclude them from selected profile-likelihood regions. We have
identified the response configurations responsible for this behavior and examined their stability under different atomic and nuclear prescriptions. These findings demonstrate that the removal of local parameter degeneracies is not
equivalent to the global determination of the underlying interaction strengths. Consequently, the number and precision of available observables alone do not guarantee that all fundamental couplings can be independently bounded.

An essential aspect of this result is its physical interpretation. The divergent directions emerging from the unrestricted analysis cannot be extrapolated indefinitely within a model that assumes fixed nuclear
sources and treats the new interaction perturbatively. To investigate this limitation, we have introduced explicit finite-coupling stress tests based on the additional nucleon--nucleon potential generated by $X_{17}$ exchange. Our numerical results show that exact reproduction of the central spectroscopic values can require sizeable modifications of the nuclear interaction, whereas solutions involving substantially smaller nucleon couplings can remain statistically competitive. This establishes that the appearance of formally unbounded directions does not imply either the physical existence of arbitrarily strong interactions or the incompatibility of $X_{17}$ with the considered measurements.
Rather, the extraction of finite individual bounds necessarily depends on the physical domain adopted for the model. The thresholds explored here serve as diagnostics of this dependence and should not be interpreted as model-independent experimental exclusions.

We have also found that the quantitative interpretation of spectroscopic tensions depends on the adopted helium measurements, nuclear response prescriptions, and assumptions concerning correlations between experimental inputs.
These dependencies emphasize the importance of treating spectroscopic observables, nuclear charge-radius extractions, and Standard Model reference quantities consistently. A more complete determination of $X_{17}$ couplings will therefore require the combined analysis of individual transition frequencies, correlated theoretical uncertainties, and nuclear responses, together with complementary particle-physics constraints. Such an approach would allow the formal parameter degeneracies identified here to be distinguished from physically admissible regions of the coupling space.

Overall, our findings establish that the sensitivity of precision spectroscopy to new physics must not be confused with its ability to uniquely determine the parameters of a new interaction. Beyond the specific $X_{17}$ scenario, this distinction is relevant to the interpretation of precision measurements in models involving multiple independently coupled particle species. The results presented here therefore provide a basis for assessing which additional measurements and theoretical developments are needed to transform increasingly accurate atomic spectroscopy into reliable constraints on physics beyond the Standard Model.

\noindent\textbf{Data and code availability.} Published input values, reproducible numerical scripts, source audits and validation outputs are included in the supplementary project archive. These scripts reproduce the compressed one-body analysis, not the primary spectroscopy datasets or a nuclear many-body calculation.


\bibliographystyle{compactnumeric}
\fontsize{9pt}{10.25pt}\selectfont
\raggedright
\raggedbottom
\interlinepenalty=10000
\bibliography{references}
\end{document}